# FPGA Implementation of RDMA-Based Data Acquisition System Over 100 GbE

Wassim Mansour, *Member, IEEE*, Nicolas Janvier, *Member, IEEE*, and Pablo Fajardo

*Abstract*— This paper presents an RDMA over Ethernet protocol used for data acquisition systems, currently under development at the ESRF. The protocol is implemented on Xilinx Ultrascale + FPGAs thanks to the 100G hard MAC IP. The proposed protocol is fairly compared with the well-known RoCE-V2 protocol using a commercial network adapter from Mellanox. Obtained results show the superiority of the proposed algorithm over RoCE-V2 in terms of data throughput. Performance tests on the 100G link show that it can reach a maximum stable link performance of 90 Gbps with minimum packets sizes greater than 1KB and 95Gbps for packet sizes greater than 32KB.

*Index Terms*— FPGA, RDMA, 100GbE, RoCE, Infiniband, data acquisition.

## I. Introduction

THE progress in manufacturing technologies and processes results in a significant increase of produced data rates in modern and upcoming 2D X-ray detectors. Such data streams are challenging to transfer, to manipulate and to process in acceptable time.

A generic and scalable data acquisition framework, called RASHPA, is currently under development at the ESRF. It will be integrated in the next generations of high performance X-ray detectors [1].

One of the key and specific features of this new framework is the use of remote direct memory access (RDMA) for fast data transfer. RDMA consists on the transfer of data from the memory of one host or device into that of another one without any CPU intervention. This allows high-throughput, low-latency networking. Companies are investing more and more into this feature, already applied to high performance computing, by integrating it into their network cards and communication adapters. Some of the available technical solutions are Infiniband [2], RDMA over Converged Ethernet (RoCE) [3] and internet Wide Area RDMA Protocol (iWARP) [4].

Paper submitted for review on June 22, 2018.
This project has received funding from the European Union's Horizon 2020 research and innovation programme under grant agreement No. 654220.
W. Mansour is with the European Synchrotron Radiation Facility, Grenoble, FRANCE; mail: wassim.mansour@esrf.fr
N. Janvier is with the European Synchrotron Radiation Facility, Grenoble, FRANCE; mail: nicolas.janvier@esrf.fr
P. Fajardo is with the European Synchrotron Radiation Facility, Grenoble, FRANCE; mail: pablo.fajardo@esrf.fr

RASHPA Framework has been prototyped and concept proven in [1] where the data link was selected to be the Peripheral Component Interconnect Express (PCIe over cable) [5]. Despite the benefits of this link, for which the native RDMA feature is the most important, it presents major limitations in terms of small transfer packet size, limited availability of PCIe over cable commercial off-the-shelf products such as switches and adapters, and the lack of standardization for optical cabling form [6].

The need to switch to a more standard networking scheme leads us to the RDMA over 100G Ethernet solution. RoCE and iWARP are two Ethernet standards in high performance computing. RoCE is a protocol developed by Mellanox and based on the Infiniband specifications. It exists in two versions: the first one is a Layer 1 protocol with an Ethernet type 8915 whereas the second one, called RRoCE (routable RoCE), is a layer 3, UDP/IP protocol, with Infiniband header inserted in the UDP data field. iWARP is another widely used RDMA over TCP/IP supported by Chelsio. A comparison between both protocols as seen from the side of Mellanox and Chelsio is presented in [7] and [8].

Both iWARP and RRoCE are heavy to be implemented on FPGA in terms of hardware resources as well as latency requirements. The first one requires a TCP/IP stack so discarded from the work performed in this paper and only RRoCE in its simplest and fastest mode called UD (Unreliable Datagram), is investigated.

The main objective of the work presented in this paper is to implement a dedicated data transfer interface over Ethernet UDP protocol together with a DMA over PCIe engine. The implementation of an ESRF RDMA over 100 Gb Ethernet solution is detailed. Two implementations should be considered, a front-end (detector transmitter side), and a back-end (computer receiver side). The front-end design is integrated within the RASHPA controller logic whereas the back-end one is supposed to be plugged into the PCIe slot of the backend computer intended to receive detector data.

The paper is organized as follows: Section II briefly introduces the concept of RASHPA. Section III provides a background and discusses the FPGA implementation challenges of RRoCE protocol. Section IV, details the proposed RDMA over Ethernet protocol. Section V experimental results as well as a comparison between the proposed RDMA protocol and RRoCE are presented. Conclusions and future perspectives are discussed in section V.

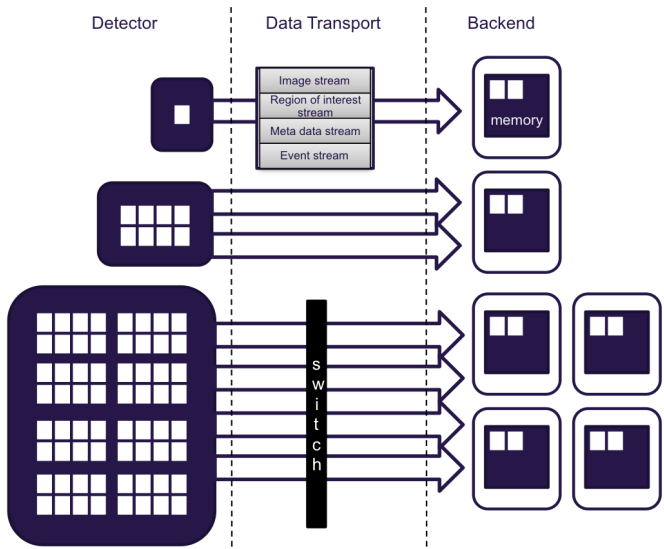

Fig. 1. Block diagram of Rashpa's concept

## II. RASHPA CONCEPT

RASHPA allows detectors to push data (images, regions of interest (ROI), metadata, events etc...) produced by 2D X-ray detectors directly into one or more backend computers. RASHPA's main properties are its scalability, flexibility and high performance. It is intended to have an adjustable bandwidth that can be compatible with any backend computer. Figure 1, shows a block diagram of different RASHPA network schemes.

In RASHPA, one can consider two types of backend computers. The first one is called System Manager (SM), and is responsible of the initialization and configuration of the whole system. The second type is called Data Receiver (DR), which is intended to receive the detector data in its local memory buffers.

The usual data destinations are random access memory buffers (RAM). Other possible destinations that are currently under investigation at the ESRF, are Graphical Processing Units (GPU), coprocessors and disk controllers.

From a hardware point of view, the RASHPA controller consists of specific logic interfacing the detector readout electronics as well as a set of hardware blocks handling data transmission. These blocks are known as channels. Two types of configurable channels can be identified in RASHPA: data and event channels.

Data channels are responsible of transferring detector data to a pre-configured address space within one or several data receivers. Multiple data channels instances can be implemented in a single RASHPA controller.

An event channel is responsible of informing the data receiver or system manager about any event occurring in the overall system. Typical events are errors, end of transmission conditions, source memory overflow etc. Only one event channel is required to be implemented a full RASHPA system.

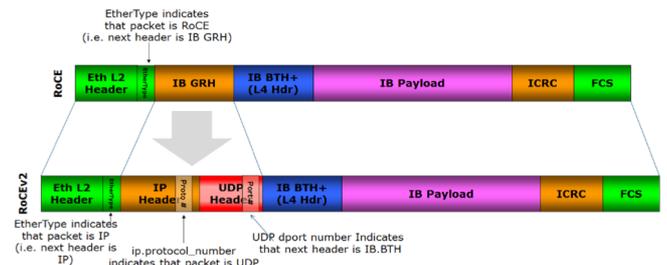

Fig. 2. RoCE-v1 vs RoCE-v2 packets

RASHPA is independent of the data link used for transmission, however a requirement that should be respected by the selected data link is the support for the RDMA feature.

## III. RDMA OVER CONVERGED ETHERNET ROCE

Ethernet is a computer networking protocol introduced in 1983 and standardized as IEEE 802.3 [9]. It divides the data stream into shorter pieces called frames. Each frame contains source and destination Media Access Controller (MAC) addresses, Ethernet type, data and error-checking code for the frame data.

The Ethernet type field specifies which protocol is to be included in the frame. Internet Protocol (IP) is one of these communication protocols and is the level 3 in the Open Systems Interconnection (OSI) model which constitute the Ethernet communication standard. User Datagram protocol (UDP) is one of the essential communication protocols used by the IP protocol. The UDP frame consists of several fields in addition to the Ethernet header and the IP header: source port, destination port, length, checksum and payload data.

RoCE (RDMA over converged Ethernet) [10] is an Ethernet protocol based on the Infiniband specification [11], and available in two different versions: RoCE-v1 and RoCE-v2 or RRoCE. RoCE-v1 is an Ethernet layer non routable protocol whereas the routable version RRoCE is the most interesting for RASHPA's implementation.

RRoCE is an RDMA capable, layer 3 network based on UDP/IPv4 or UDP/IPv6, and relying on congestion control and lossless Ethernet. It is currently supported by several off-the-shelf network adapters as well as the latest Linux kernel drivers.

The UDP payload data of a RRoCE protocol, illustrated in figure 2, contains an Infiniband header, the actual data payload in addition to an invariant cyclic redundancy check (iCRC) field that is mandatory for the RoCE packets in order to be accepted by the network adapter. The iCRC field is retained from the Infiniband specifications. Figure 3, show the iCRC32 calculation algorithm. Note that, an

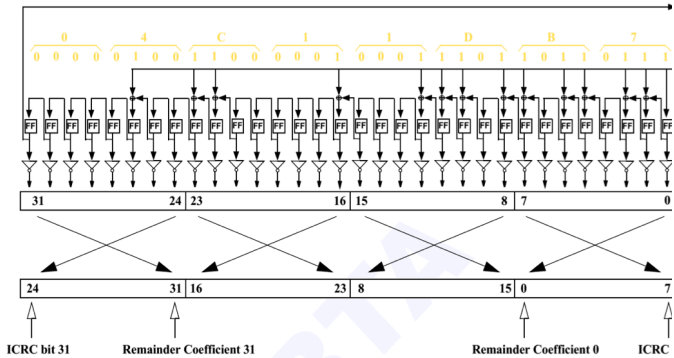

Fig. 3. Calculation of invariant CRC for RoCE protocol

Ethernet frame does also contain another CRC field for the global packet.

The calculation of the iCRC algorithm for RoCE-V2/IPv4 is performed following the below steps:

1) Extract RoCEv2: IP+UDP+InfiniBand.

2) Add Dummy LRH field, 64 bits of 1's. This field is present in the Infiniband specifications, so in order to have correct CRC calculation one have to include its dummy bits.

3) For RoCEv2 over IPv4

    Time to Live = 1's

    Header Checksum = 1's

    Type of Service (DSCP and ECN) = 1's

4) UDP checksum = 1's.

5) Resv8a field on Infiniband protocol = 1's

7) CRC calculation is based on the crc32 used for Ethernet networking, 0x04C11DB7.

8) CRC calculation is done over the UDP frame starting from the most significant bit of the most significant byte.

9) Inversion and byte swap has to be applied in order to get the invariant arc to be integrated in the RRoCE frame.

A first FPGA implementation trial of the RRoCE has been performed using the unreliable datagram mode (UD) [3]. In this mode data are sent in streams without any acknowledgement from the receiver side. The target FPGA board was the KCU116 by Xilinx [12] and the target network adapter was a Mellanox ConnectX-4 (MCX415A-CCAT) board. It is important to note that in Ultrascale+ families, the 100G CMAC IP core is a hard IP having LBUS (Local BUS) as input/output, which have to be converted into AXI stream bus to be integrated in system on chip designs.

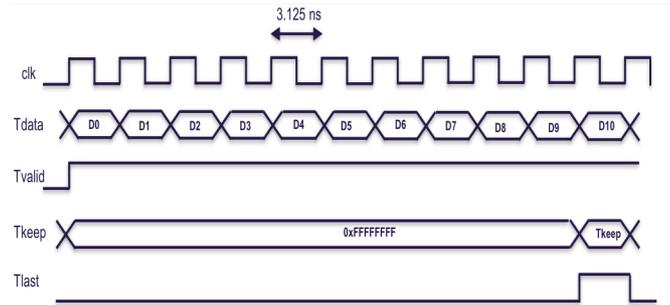

Fig. 4. Timing diagram of the AXI stream data used for the iCRC calculation

In fact, the basic challenge in the FPGA implementation of RRoCE algorithm is the optimal implementation of an iCRC algorithm. Figure 4 depicts the timing diagram of the input stream data used for the iCRC calculation. Data of 64 bytes are streamed at each 3.125 ns clock cycle period except the last cycle that may contain partial data that requires multiplexing via the AXI stream "tkeep" signal for byte selection.

A pipelined iCRC design requires 64 clock cycles in order to calculate the iCRC over the 64-bytes input. After 64 clock cycles, the design will be allowed to continue the calculation over the second 64-bits input data. That means that 200 ns are lost for each data calculation of 64 bytes. Supposing that the transmitter sends 12.5GB (100 Gbits) of data, that will theoretically take one second to be transferred over a 100Gbps Ethernet link, the actual theoretical transfer delay caused by the iCRC calculation will be 42 ms that is 4.2%.

IV. THE PROPOSED RDMA OVER RDMA PROTOCOL

RRoCE is a well-developed commercial protocol supported by the ib-verbs library available in the latest Linux kernels. However, one can even go faster in data transfer due to the iCRC calculation problem and the overhead used for the Infiniband header. In addition to the previously mentioned reasons, controllability and observability over an in-house developed protocol is a major advantage for an ESRF RDMA over Ethernet protocol over RoCE.

The proposed RDMA over Ethernet standard proposed in this paper will mainly use the UDP/IP protocol for routability, and information about each transfer in the unused source and destination ports of the UDP header.

The proposed standard relies on the interactions of three major actors. The first one is the RASHPA controller on the X-ray detector front-end side which is the data transmitter. The second one is the FPGA board acting as a data receiver, that will transform UDP packets coming from the transmitter into PCIe DMA-based packets. These packets are sent to some buffers on the data receiver computer which is the third actor in the system. Figure 5 illustrates the architecture of the overall system.

There will be a software library called LIBRASHPA

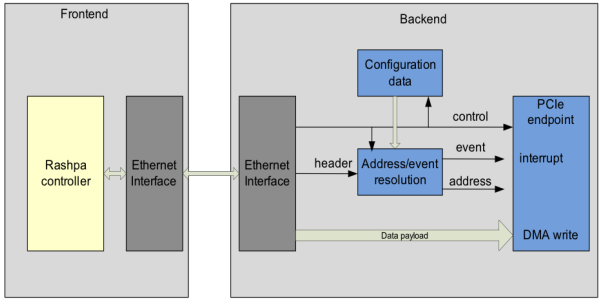

Fig. 5. Architecture of the proposed RDMA over Ethernet protocol

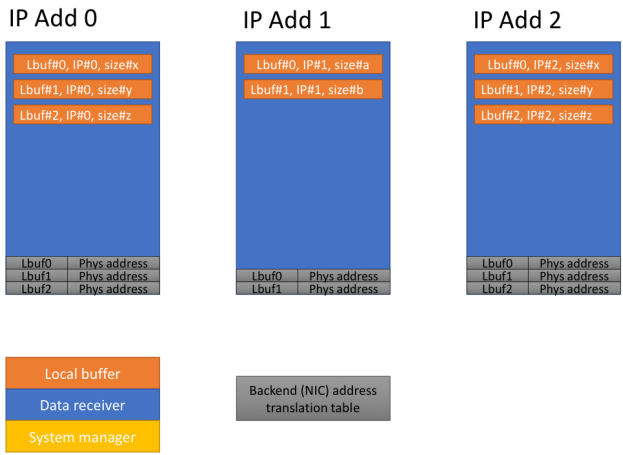

Fig. 6. Representation of local memory buffer in the backend computers

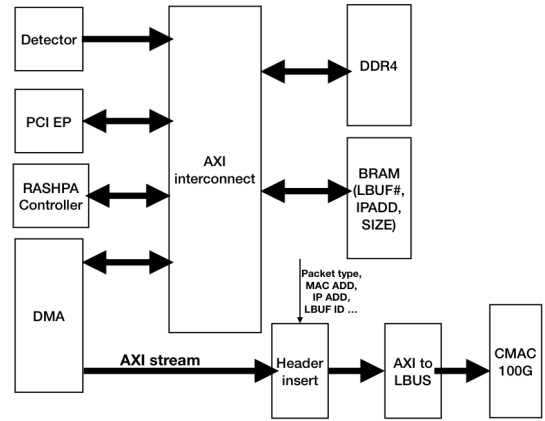

Fig. 7. FPGA implementation of the RDMA over 100G transmitter side

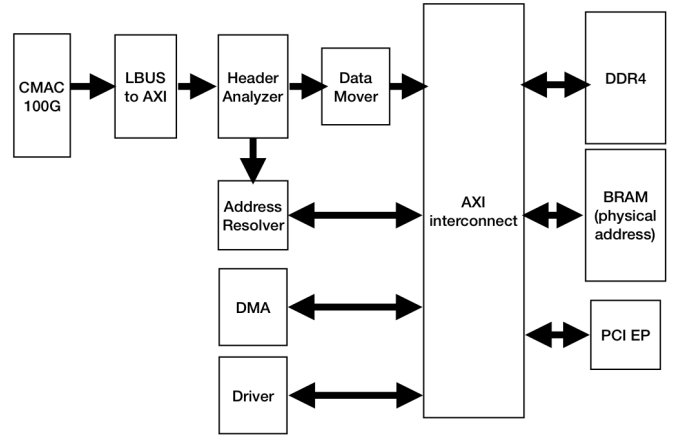

Fig. 8. FPGA implementation of the RDMA over 100G Ethernet receiver side

installed on the data receiver side that will help allocating memory buffers of different sizes to be used as final data destinations. These buffers will be identified by an identification number (ID), a size, and the IP address of the data receiver as depicted in figure 6. The RASHPA controller, which is the transmitter, should have enough knowledge about these three parameters, however the receiver FPGA board should store the real physical address of the allocated buffers for address translation.

Figure 7, shows the FPGA implementation of the Ethernet transmitter side using the Xilinx 100G cmac IP. Data streams coming from the detector are stored in a DDR4 memory. Whenever a full image is written to the DDR, the RASHPA controller will configure a Direct Memory Access (DMA) IP allowing it to read the data via an AXI4 interconnect, and sends it as stream of data (AXI stream bus) to the header insertion IP. The header insertion IP gets its configuration from the RASHPA controller. In fact, the configuration of the header insertion unit is nothing but the UDP header and the destination local buffer represented by the identification parameters stored at the initialization phase in an internal Block RAM (BRAM). The constituted header will be concatenated with the data stream coming from the DMA. Since the CMAC IP has a local bus (LBUS) input/output interface, a bridge between the AXIS to LBUS has been implemented and used as an intermediate stage between the header insertion unit and the CMAC IP. The configuration of the whole process can be done using the same Ethernet link or via an external link such as 1Gb Ethernet, PCIe over cable, etc.

At the receiver side, figure 8, the CMAC output data as LBUS are bridged to an AXI stream interface before it gets analysed in order to resolve the physical address of the final destination buffer. Actually, during the initialization phase, LIBRASHPA should store the physical address of each local buffer in a BRAM inside the receiver's FPGA. The output data of the header analyser unit can be stored in a DDR4 or FIFO for synchronization, then sent to the PCI express endpoint for DMA transfer to the final destination. The whole process is controlled by a finite state machine implemented in the driver IP.

In order guarantee the no packet loss, one can use a converged network, but in case of lost packets, the data receiver should be informed. For that, a simple packet loss detection algorithm has been implemented. It consists of a 1024-bit shift register. Each bit in this shift register represents one packet number represented by its sequence number. When packet sequence number "512" is received, the receiver checks packet "1", if it is missing, than it generates an event to inform the data receiver. The same process repeats for each received packet, which means that the receiver can identify a lost packet after 512 received packets. The process is illustrated in figure 9.

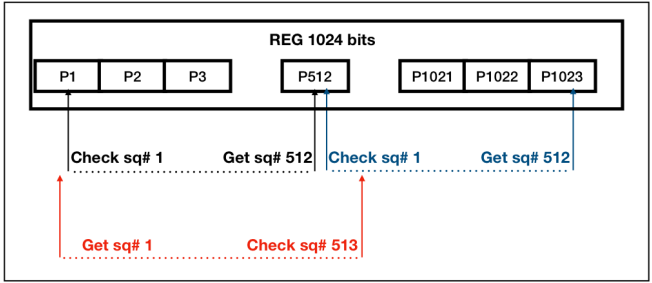

Fig. 9. Packet loss detection algorithm

## V. EXPERIMENTAL RESULTS

The implementation of the proposed prototype as well as RoCE-V2, at the transmitter side, targets a Xilinx FPGA development board (KCU116). The board is based on the XCKU5P Kintex Ultrascale+ family. In case of the proposed prototype, the receiver implementation targets an industrial board called XpressVUP developed by Reflexces[13]. It is based on a XCVU9P virtex ultrascale+ FPGA with an integrated Gen3x16 PCIe endpoint. The PCIe endpoint is comparable to the integrated one in the Mellanox network adapter card, MCX415A-CCAT, used as a RoCEV2 backend. A UDP stack has been implemented on the transmitter FPGA allowing the RASHPA controller to construct frames of data and the back-end to read these packets and analyse them before transforming them into DMA configurations. Post route of the front-end (transmitter) FPGA implementation show that the design occupies around 50% of the total CLBs and 21 % of BRAM of the selected XCKU5P FPGA.

To confirm the correctness of the constructed packets and to test the transfer bandwidth, the Mellanox NIC was used together with wireshark software on a PC running on Linux debian distribution.

The realized experiments allow building correct UDP packets, however the UDP receive buffer overloaded when measuring UDP bandwidth due to the high transfer rate without the ability to empty it. Hardware RoCE-V2 as well as soft-RoCE were also tested between two mellanox boards running at 100Gbps.

In order to provide a fair comparison of the transfer throughput of both protocols, one should exclude the CRC implementation because it will terribly affect the transfer rate.

First of all, and in order to have an idea about the transfer one could achieve with the 100G link itself, FPGA to FPGA UDP transfers were selected. Different configurations of the MAC IP including packet sizes and number of packets to send were selected. Figure 10 illustrates the obtained results, and shows that the 100G transfer can reach a rate of 90Gbps for a minimum packet sizes of 1KB and becomes stable at 95Gbps for packet sizes of 32KB and above. Small packet sizes decrease significantly the throughput

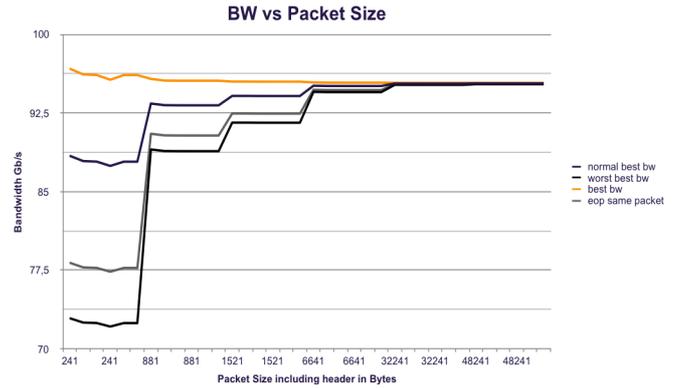

Fig. 10. Bandwidth Vs Packet size for the 100GbE link

The throughput comparison between RoCE and the proposed algorithm was based on pre-constructed data packets of 598 bytes. The same configuration was adapted for both algorithms where a computer was used to configure the DMA on the transmitter side for each transfer. Note that this is not the optimal throughput to measure because of the CPU interaction at each packet.

TABLE I
RoCE-V2 VS THE PROPOSED PROTOYPE MEASURED BANDWIDTH

| Protocol | Data Transfer | Bandwidth | |
|---|---|---|---|
| RoCE-V2 | 598Bytes | 6.1 Gbps | 1 |
| Proposed protocol | 598Bytes | 10.3 Gbps | 1 |

Table I, presents the measured bandwidth for both algorithms using the adopted strategy.

Results show that the proposed algorithm is more than 1.5 times faster than the RoCE-V2 protocol considering that the iCRC is pre-calculated and only the link is tested together with the receiver side, i.e the Mellanox network adapter versus the FPGA implementation of the suggested protocol. Both receivers are connected via PCIe x16 lanes.

It is important to note that while performing these end-to-end tests, either from one FPGA to another or from an FPGA to Mellanox board, no lost packets were detected.

## VI. CONCLUSIONS AND FUTURE WORK

This paper presented a dedicated data transfer protocol based on remote direct memory access over Ethernet. The protocol is intended to be used in the next detector generations that are under development at the ESRF. The implementation was realized on a KCU116 xilinx development board and compared with the commercial widely used protocol RoCE-V2 implemented on the same FPGA board and wired to a Mellanox network adapter connect-X4 board.

Comparison results show the superiority in terms of data throughput of the proposed protocol with respect to RRoCE even when excluding the iCRC calculation.

Future development will focus on the integration of both the proposed protocol and RoCE all together in the RASHPA framework. Selection between these protocols will be based on the price/throughput requirements for each detector application.

Testing the protocol over a routable network of detectors/backend computers is the next goal of the project.

REFERENCES

[1] W. Mansour et al., "High performance RDMA-based DAQ platform over PCIe routable network", presented at ICALEPCS, Barcelona, Spain, Oct. 8-13, 2017.
[2] Mellanox Technologies, Introduction to Infiniband, White paper, Document Number 2003WP.
[3] Mellanox Technologies, "RDMA/ROCE solutions", [Online] Available: https://community.mellanox.com/docs/DOC-2283.
[4] Intel Corp., "Understanding iWARP: Delivering low latency Ethernet", [Online] Available: https://www.intel.fr/content/www/fr/fr/ethernet-products/gigabit-server-adapters/iwarp-brief.html
[5] PCI-SIG, PCI Express base specification revision 3.0, November 10, 2010.
[6] F. Le Mentec et al., "RASHPA: A data acquisition framework for 2D X-Ray detectors", presented at ICALEPCS, San Francisco, United States, Oct. 6-11, 2013.
[7] Mellanox Technologies, RoCE Vs iWARP, White paper, [online]. Available: http://www.mellanox.com/related-docs/whitepapers/WP_RoCE_vs_iWARP.pdf
[8] Chelsio, RoCE FAQ, [online]. Available: https://www.chelsio.com/roce/
[9] IEEE Ethernet standard, IEEE Standard 802.3, 1983.
[10] RoCEV2, Supplement to Infiniband architecture specification volume 1, Annex A17, Release 1.2.1, 2014
[11] Infiniband architecture specification Volume 1, Release 1.2.1, November 2007
[12] Xilinx Development Board, [Online]. Available : https://www.xilinx.com/products/boards-and-kits/ek-u1-kcu116-g.html
[13] XpressVUP-LP9P, [Online]. Available : https://www.reflexces.com/products-solutions/other-cots-boards/xilinx/xpressvup